\documentclass{emulateapj} \newcommand{\ltab}{\LongTables} \newcommand{\rotate}{}

\usepackage{hyperref}

\hypersetup{
    bookmarks=true,         
    unicode=false,          
    pdftoolbar=true,        
    pdfmenubar=true,        
    pdffitwindow=true,     
    pdfstartview={FitH},    
    pdftitle={Magellan series II},    
    pdfauthor={vmplacco},     
    pdfsubject={Astronomy},   
    pdfcreator={dvipdf},   
    pdfproducer={dvipdf}, 
    pdfkeywords={metal-poor stars},
    pdfnewwindow=true,      
    colorlinks=true,       
    linkcolor=red,          
    citecolor=blue,        
    filecolor=magenta,      
    urlcolor=cyan,           
    breaklinks=true,
    linktocpage
}

\newcommand{\metal}{[Fe/{}H]}
\newcommand{\cfe}{[C/{}Fe]}
\newcommand{\abund}[2]{[#1/{}#2]}

\newcommand{\teff}{$T_{\rm eff}$\,}
\newcommand{\logg}{log\,$g$\,}

\newcommand{\jk}{({$J-K$})$_0$}
\newcommand{\xx}{{\tablenotemark{a}}}

\shorttitle{Metal-Poor Stars Observed with the Magellan Telescope. II.}
\shortauthors{Placco et al.}

\begin{document}

\title{Metal-Poor Stars Observed with the Magellan Telescope. II.\footnotemark[1]\\
Discovery of Four Stars with \metal~$\le -3.5$}

\author{Vinicius M. Placco\altaffilmark{2,3},
Anna Frebel\altaffilmark{4},
Timothy C. Beers\altaffilmark{2},
Norbert Christlieb\altaffilmark{5},\\
Young Sun Lee\altaffilmark{6},
Catherine R. Kennedy\altaffilmark{7},
Silvia Rossi\altaffilmark{3},
Rafael M. Santucci\altaffilmark{3}}

\altaffiltext{1}{Based on observations gathered with: 
the 6.5 meter Magellan Telescopes 
located at Las Campanas Observatory, Chile;
Southern Astrophysical Research (SOAR) telescope (SO2011B-002), 
which is a joint project of the Minist\'{e}rio da Ci\^{e}ncia, Tecnologia, e 
Inova\c{c}\~{a}o (MCTI) da Rep\'{u}blica Federativa do Brasil, the U.S. 
National Optical Astronomy Observatory (NOAO), the University of North Carolina at 
Chapel Hill (UNC), and Michigan State University (MSU); and New Technology Telescope 
(NTT) of the European Southern Observatory (088.D-0344A), La Silla, Chile.}

\altaffiltext{2}{National Optical Astronomy Observatory, Tucson, AZ 85719, USA}

\altaffiltext{3}{Departamento de Astronomia - Instituto de Astronomia, 
Geof\'isica e Ci\^encias Atmosf\'ericas, Universidade de S\~ao Paulo, 
S\~ao Paulo, SP 05508-900, Brazil}

\altaffiltext{4}{Massachusetts Institute of Technology, Kavli Institute for
  Astrophysics and Space Research, 77 Massachusetts Avenue, Cambridge,
  MA 02139, USA}
  
\altaffiltext{5}{Zentrum f\"ur Astronomie der Universit\"at Heidelberg, 
Landessternwarte, K\"onigstuhl 12, 69117, Heidelberg, Germany}
  
\altaffiltext{6}{Department of Astronomy, New Mexico State University, 
Las Cruces, NM 88003, USA}

\altaffiltext{7}{Research School of Astronomy and Astrophysics, Australian 
National University, Canberra, ACT 2611, Australia}

\begin{abstract}

We report on the discovery of seven low-metallicity stars selected from
the Hamburg/ESO Survey, six of which are extremely metal-poor
(\metal~$\leq-$3.0), with four having \metal~$\leq-$3.5. Chemical
abundances or upper limits are derived for these stars based on
high-resolution (R $\sim~35,000$) Magellan/MIKE spectroscopy, and are in
general agreement with those of other very and extremely metal-poor
stars reported in the literature. Accurate metallicities and abundance
patterns for stars in this metallicity range are of particular
importance for studies of the shape of the metallicity distribution
function of the Milky Way's halo system, in particular for probing the
nature of its low-metallicity tail. In addition, taking into account
suggested evolutionary mixing effects, we find that six of the program
stars (with \metal~$\leq-$3.35) possess atmospheres that were likely originally
enriched in carbon, relative to iron, during their main-sequence phases.
These stars do not exhibit over-abundances of their $s$-process elements,
and hence may be additional examples of the so-called CEMP-no class of
objects. 

\end{abstract}


\section{Introduction}
\label{intro}

The very metal-poor (VMP; [Fe/H]\footnote{\abund{A}{B} = $log(N_A/{}N_B)
_{\star} - log(N_A/{}N_B) _{\odot}$, where $N$ is the number density of
atoms for a given element, and the indices refer to the star ($\star$)
and the Sun ($\odot$).} $\leq -2.0$) and extremely metal-poor (EMP;
[Fe/H] $\leq -3.0$) stars provide a direct view of Galactic chemical and
dynamical evolution; detailed spectroscopic studies of these objects are
the best way to identify and distinguish between a number of possible
scenarios for the enrichment of early star-forming gas clouds soon after
the Big Bang. Thus, over the last 25 years, several large survey efforts
were carried out in order to dramatically increase the numbers of VMP
and EMP stars known in our galaxy, enabling their further study with
high-resolution spectroscopy. Taken together, the early HK survey
\citep{beers1985,beers1992} and the Hamburg/ESO Survey
\citep[HES;][]{wisotzki1996,christlieb2001,christlieb2003,christlieb2008} 
identified several thousand VMP stars (Beers \& Christlieb 2005; for a
more recent summary of progress, see Frebel \& Norris 2011). To date,
the two most metal-poor (strictly speaking, most iron-deficient) stars
found in the halo of the Galaxy, HE~0107$-$5240 \citep[\metal\ = $-$5.2;
][]{christlieb2002} and HE~1327$-$2326 \citep[\metal\ = $-$5.6;
][]{frebel2005}, were first identified from spectroscopic follow-up of
candidate VMP/EMP stars in the HES stellar database. 

In the past decade, the numbers of metal-poor stars has been
further increased, by over an order of magnitude, through the use of
medium-resolution spectroscopy carried out during the Sloan Digital Sky
Survey \citep[SDSS;][]{york2000} and the sub-surveys Sloan Extension for
Galactic Understanding and Exploration \citep[SEGUE-1;][]{yanny2009} and
SEGUE-2 (C. Rockosi et al., in preparation), to many tens of thousands of
VMP (and on the order of 1000 EMP) stars. Although follow-up
high-resolution spectroscopy of likely VMP/EMP stars from SDSS/SEGUE has
only recently begun \citep[e.g.,][]{caffau2011a,bonifacio2012,aoki2013},
the SDSS/SEGUE sample has already resulted in the discovery of several
stars in the ultra metal-poor (UMP; \metal~$< -4.0$) range,
SDSS~J102915$+$172927, with \metal\ = $-$4.73 \citep{caffau2011b},
SDSS~J144256$+$253135, with \metal\ = $-$4.09 \citep{caffau2013}, and
another star, SDSS J174259$+$253135, for which iron lines could not be
measured, but with [Ca/H] $< -4.5$ \citep{caffau2013}.

There exists a substantial number of published high-resolution studies
of stars identified by the above surveys (and others), due to the wealth
of information concerning early element production that can be
extracted. Examples of these efforts are the ``Survey of Proper Motion
Stars'' series \citep[e.g.,][]{carney1994}, the ``First Stars'' project
\citep[][among others in the series]{cayrel2004, francois2007},
the ``The 0Z Project'' \citep{cohen2008,cohen2011,cohen2013}, the ``Extremely
Metal-Poor Stars from SDSS/SEGUE'' series \citep{aoki2013}, and ``The
Most Metal-Poor Stars'' series \citep{norris2013a, yong2013}, as well as
smaller individual samples from \citet{ryan1991,ryan1996,ryan1999},
\citet{mcw1995}, \citet{lai2008}, \citet{hollek2011}, and others.
\citet{roederer2013} describe a high-resolution spectroscopic
analysis of some 300 VMP/EMP stars originally identified by the HK
Survey. These studies all aim to provide large, homogeneous datasets of
metal-poor stars, and to address differences in their chemical abundance
patterns at the lowest metallicities. Taking into account the published
work that we are aware of, the current total number of EMP stars with
available high-resolution spectroscopic determinations is 205. This
number drops to 50 for \metal~$<-3.5$, 12 for \metal~$<-4.0$, and only 4
stars observed to date with \metal~$<-4.5$ \citep{saga2008, frebel10,
yong2013}. 

The number of EMP stars with well-determined metallicities also has a
direct impact on the observed metallicity distribution function (MDF)
for the halo system of the Milky Way. The MDF provides important
constraints on the structure and hierarchical assembly history of the
Galaxy \citep[see, e.g.,][]{carollo2007,carollo2010,tissera2010,
beers2012,mccarthy2012,tissera2012,tissera2013}, on quantities such as
the initial mass function and early star-formation rate 
\citep[see, e.g.,][]{romano2005,pols2012,chiappini2013,lee2013,suda2013},
and on tests of models for Galactic chemical evolution 
\citep[see e.g.,][and references therein]{romano2005,chiappini2013,karlsson2013,nomoto2013}. 

Recent studies based on observations of metal-poor giants and
main-sequence turnoff stars from the HES \citep{schorck2009, li2010}
suggest that the halo MDF has a relatively sharp cutoff at
\metal~$\sim-3.6$, along with a handful of UMP and hyper metal-poor
(HMP; \metal~$< -5.0$) stars. It is important to understand whether this
claimed shortfall in the expected numbers of stars at the lowest
metallicities is real, and perhaps the result of the nature of the
progenitor mini-halos that contributed to the assembly of the Galactic
halo system, or simply an artifact introduced by, e.g., subtle (or not
so subtle) biases in observational follow-up strategies or 
target-selection criteria. The high-resolution spectroscopic study of
\citet{yong2013b} addresses precisely this question. Based on a sample
of 86 stars with \metal\ $< -$3.0, and 32 stars with \metal\ $<-$3.5,
determined from uniformly high-S/N spectra analysed in a homogeneous
fashion, the authors find no evidence to support a sharp cutoff in the
halo system's MDF at \metal\ = $-$3.6, as suggested by \citet{schorck2009}
and \citet{li2010}. In fact, based on their data, Yong et al. state that
``the MDF decreases smoothly down to \metal$=-$4.1''. It is clear that
this matter is still under discussion, hence additional EMP stars, in
particular those with \metal~$<-3.5$, are required to better assess the
true nature of the halo system's MDF, in particular its low-metallicity
tail.

As one explores the metallicity regime below [Fe/H] = $-$3.0, it is
expected that the chemically-primative low-mass stars formed in the
early history of the Galaxy preserve in their atmospheres the
``elemental abundance fingerprints'' of the first few stellar
generations, the stars responsible for the first nucleosynthesis
processes to occur in the Galaxy. In order to yield the abundance
patterns observed today in EMP stars, these first-generation objects
should have evolved quickly, and hence are expected to have been
associated with stars of high mass. Indeed, present simulations of the
formation of the first stars \citep[see][for a recent review]{bromm2011}
generally point toward masses in the range 30 to 100 $M_\odot$, the
so-called Pop III.1 stars, although some have also explored lower mass
ranges, including characteristic masses on the order of 10 $M_\odot$,
the Pop III.2 stars, or even lower. 

One of the important enrichment scenarios for EMP stars is based on
material ejected by first-generation, core-collapse supernovae (SNe) in
the early universe that, under certain circumstances (likely involving
rapid rotation), undergo a process referred to as ``mixing and fallback,
'' \citep[e.g.,][]{nomoto2006,nomoto2013}; the mass range considered
for these stars is 25 to 50 $M_\odot$, but it is still rather uncertain.
There are also models for more massive, very rapidly-rotating, mega
metal-poor (MMP; [Fe/H] $< -6.0$) stars \citep{meynet2006,meynet2010}.
Both of these sources may be important contributors to the abundance
patterns observed among many EMP stars. Details of the progenitors, such
as their mass range, rotation range, binarity, etc., might be
responsible for the differences in the observed abundance patterns among
individual EMP stars. 

It has been recognized since the early work by \citet{rossi1999},
\citet{marsteller2005}, and \citet{lucatello2006}, that a large fraction of EMP stars
\citep[between 30\% and 40\%; see][]{placco2010,lee2013} present carbon
over-abundances relative to iron \citep[using the criterion \cfe\ $> +$0.7
suggested by][]{aoki2007}. The elemental abundance patterns of these
carbon-enhanced metal-poor \citep[CEMP - originally defined
by][]{beers2005} stars can help probe the nature of different progenitor
populations, such as those mentioned above. Moreover, recent studies
\citep{aoki2007,norris2013b} show that the majority of CEMP stars with
\metal~$<-3.0$ belong to the CEMP-no sub-class, characterized by the
lack of strong enhancements in neutron-capture elements
\citep[\abund{Ba}{Fe} $<$ 0.0; ][]{beers2005}. The brightest
EMP star in the sky, BD+44:493, with [Fe/H] $= -3.8$ and $V = 9.0$, is a
CEMP-no star \citep{ito2009,ito2013}. The distinctive elemental abundance
pattern associated with CEMP-no stars (which includes enhancements of
light elements in addition to C, such as N, O, Na, Mg, Al, and Si) has been
identified in high-$z$ damped Lyman-alpha systems \citep{cooke2011,cooke2012},
and has also been found among stars in the so-called ultra-faint dwarf 
spheroidal galaxies \citep[e.g.,][]{norris2010}. 

The main goal of this work is to seek further constraints on the
nucleosynthetic processes that took place at early times in the
formation and evolution of our Galaxy. To accomplish this, we consider
seven newly discovered VMP/EMP stars identified from medium-resolution
spectroscopic follow-up of metal-poor candidates selected from the HES,
including six EMP stars that we tentatively classify as CEMP-no stars,
based on high-resolution spectroscopic analysis. We have also determined
a number of elemental abundances for our program stars, including
carbon, nitrogen, and the neutron-capture elements, which are useful for
further distinguishing between different enrichment scenarios for
these stars. 

This paper is outlined as follows. Section~\ref{secobs} presents details
of the target selection, as well as the medium- and high-resolution
spectroscopic observations carried out for the program stars. The
determination of the stellar parameters and comparison with
medium-resolution estimates of these parameters are presented in
Section~\ref{params}, followed by a detailed abundance analysis in
Section~\ref{abundsec}. Comparison between the abundances of the
observed targets and stars from other high-resolution spectroscopic
studies, together with astrophysical interpretations and possible
formation scenarios for our program stars, are presented in
Section~\ref{discuss}. Finally, our conclusions and perspectives for
future work are given in Section~\ref{final}.

\section{Target Selection and Observations}
\label{secobs}

Our program stars were originally selected according to the strength of
their Ca\,{\sc{ii}} K lines in low-resolution (R $\sim$ 300) objective-prism
spectra, in comparison with their broadband de-reddened $(J-K)_0$
colors. Medium-resolution (R $\sim$ 2,000) spectroscopic observations were
then carried out in order to determine first-pass estimates of their
stellar atmospheric parameters. Finally, high-resolution (R $\sim$ 35,000)
spectra were gathered, in order to determine elemental abundances (or upper
limits) for as many elements as possible, as described below.

\subsection{Low-Resolution Spectroscopy}

In this work, we continued the search for metal-poor stars from the HES
stellar database. This search, which used procedures similar to those
described in \citet{frebel2006} and \citet{christlieb2008}, was based on the KPHES
index \citep[which measures the Ca\,{\sc{ii}} K line strength, as
defined by][]{christlieb2008}, and the \jk{} color from 2MASS \citep{skrutskie2006}. 
The reddening corrections were determined based on the \citet{schlegel1998} dust maps. 
This search criteria led to a subsample of
metal-poor candidates (including bright sources) with KPHES $<$ 5.0
{\AA} (which removes stars with strong Ca\,{\sc{ii}} lines, regardless of
their effective temperature), and in the color range 0.50 $ \leq$ \jk $\leq$
0.70. It is important to note that the stars analyzed in this work were
also independently selected as metal-poor candidates by
\citet{christlieb2008}.

\subsection{Medium-Resolution Spectroscopy}

Medium-resolution spectroscopic follow-up was carried out along with
observations of the sample of CEMP candidates described by
\citet{placco2011}, using the 4.1m SOAR Telescope and the 3.5m ESO New
Technology Telescope (NTT). This is a necessary step in order to
determine reliable first-pass stellar parameters, before obtaining
high-resolution spectroscopy of the most promising stars.

Most of the observations were carried out in semester 2011B, using the
Goodman Spectrograph on the SOAR telescope, with the
600~l~mm$^{\rm{-1}}$ grating, the blue setting, a 1$\farcs$03 slit, and
covering the wavelength range 3550-5500\,{\AA}. This combination yielded
a resolving power of $R\sim 1500$, and signal-to-noise ratios S/N$\sim
30$ per pixel at 4000\,{\AA} (using integration times between 15 and 30
minutes). NTT/EFOSC-2 data were also gathered in semester 2011B with a
similar setup, using Grism\#7 (600~gr~mm$^{\rm{-1}}$) with a 1$\farcs$00
slit, covering the wavelength range of 3400-5100\,{\AA}. The resolving
power ($R\sim 2000$) and signal-to-noise ratios S/N $\sim 40$ per pixel
at 4000\,{\AA} (using integration times between 2 and 12 minutes), were
similar to the SOAR data. 

\begin{figure}[!ht]
\epsscale{1.20}
\plotone{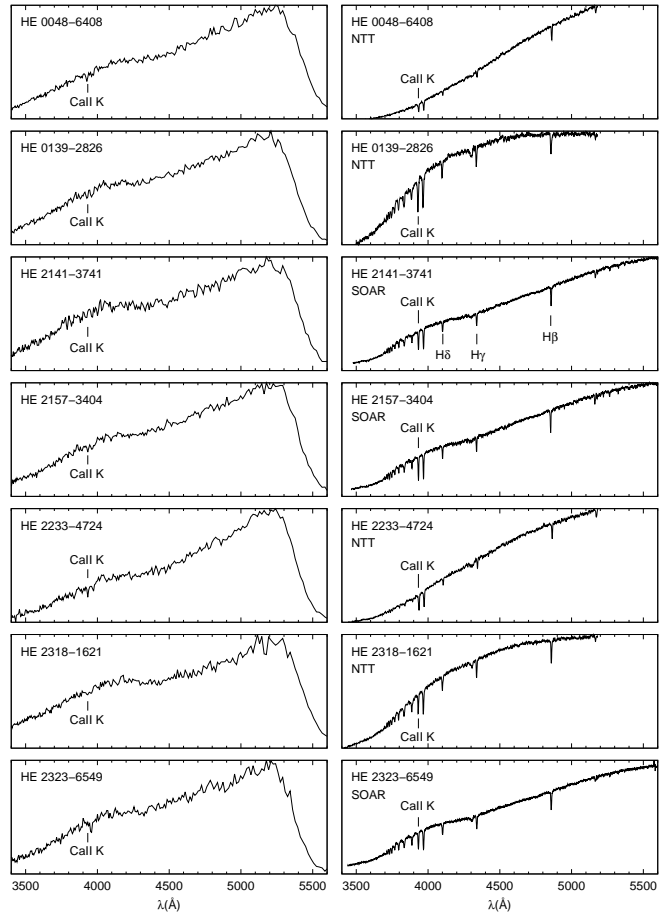}
\caption{Comparison between low-resolution HES spectra (left panels) and 
SOAR/NTT medium-resolution spectra (right panels). Prominent spectral
lines are identified in the medium-resolution specta.}
\label{low_med}
\end{figure}

The calibration frames in both cases included HgAr and Cu arc-lamp
exposures (taken following each science observation), bias frames, and
quartz-lamp flatfields. All tasks related to calibration and spectral extraction
were performed using standard IRAF\footnote{\href{http://iraf.noao.edu}
{http://iraf.noao.edu}.} packages. Table~\ref{candlist}
lists the details of the medium-resolution observations for each star.
Figure~\ref{low_med} shows the comparison between the low-resolution HES
objective-prism spectra and the medium-resolution spectra of the seven
program stars. One can note the lack of measurable features in the HES
spectra, which is an indication of the low metallicity of the targets.
In addition, at the resolution of the prism plates, there are no strong
hydrogen lines visible from the Balmer series, suggesting that the
targets have cooler effective temperatures. However, as can be seen from
inspection of the medium-resolution spectra, it is possible to identify
the Ca\, {\sc ii} lines, as well as a few hydrogen lines from the Balmer
series, as labeled in Figure~\ref{low_med}.

\subsection{High-Resolution Spectroscopy}

\begin{figure}[!ht]
\epsscale{1.20}
\plotone{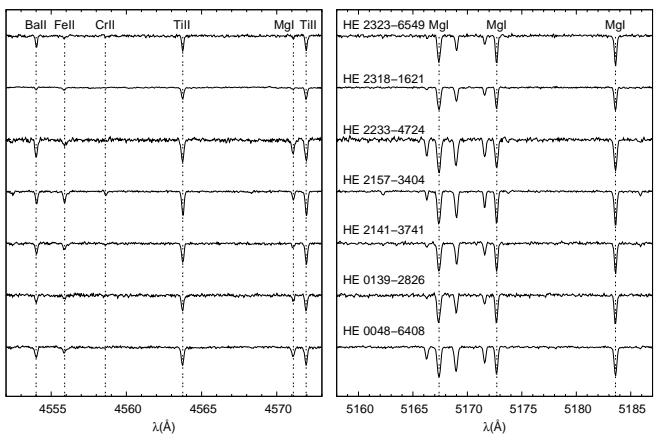}
\caption{Examples of high-resolution spectra for the program stars
in the region around the Ba 4554\,{\AA} line (left panels) and the
\ion{Mg}{1} triplet (right panels).}
\label{ba_mg}
\end{figure}

High-resolution data were obtained, during semesters 2011B, 2012B, and 2013B,
using the Magellan Inamori Kyocera Echelle spectrograph \citep[MIKE
--][]{mike} on the Magellan-Clay Telescope at Las Campanas Observatory.
The observing setup included a 0$\farcs$7 and a 1$\farcs$0 slits with 2$\times$2 on-chip
binning, which yielded a resolving power of R~$\sim35,000$ in the blue
spectral range and R~$\sim28,000$ in the red spectral range, with an
average S/N $\sim85$ per pixel at 5200\,{\AA} (using integration times between 15
and 60 minutes). MIKE spectra have nearly full optical wavelength
coverage from $\sim$3500-9000\,{\AA}. Table~\ref{candlist} lists the
details of the high-resolution observations for the program stars. These
data were reduced using a data reduction pipeline developed for MIKE
spectra \footnote{http://code.obs.carnegiescience.edu/python}.
Figure~\ref{ba_mg} displays regions of the MIKE spectra -- the
left panel shows the region around the Ba\,{\sc ii} line at 4554\,{\AA}; the right panel
shows the Mg triplet around 5170\,{\AA}.

\subsection{Line Measurements}

Measurements of atomic absorption lines were based on a line list
assembled from the compilations of \citet{aoki2002} and
\citet{barklem2005}, as well as from a collection retrieved from the
VALD database \citep{vald}. References for the atomic $gf$ values can be
found in these references. Equivalent-width measurements were obtained by
fitting Gaussian profiles to the observed atomic lines. Table~\ref{eqw}
lists the lines used in this work, their measured equivalent widths, and
the derived abundances from each line. Lines with equivalent widths
marked as ``syn'' refer to abundances calculated by spectral synthesis
(see Section~\ref{abundsec} for details).

\section{Stellar Parameters}
\label{params}

Estimates of the stellar atmospheric parameters from medium-resolution
SOAR/NTT spectra were obtained using the n-SSPP, a modified version of
the SEGUE Stellar Parameter Pipeline \citep[SSPP - see][for a 
description of the procedures used] {lee2008a,lee2008b,allende2008,
smolin2011,lee2011,lee2013}. Details of the implementation of these
routines for use with non-SDSS spectra can be found in T. Beers et al. (in
preparation). Table~\ref{obstable} lists the calculated \teff,
\logg, and \metal, used as first-pass estimates of parameters for the
high-resolution analysis.

From the high-resolution spectra, effective temperatures of the stars
were determined by minimizing trends between the abundance of Fe\,{\sc
i} lines and their excitation potentials. This procedure is known to
underestimate the effective temperatures relative to those determined
based on photometry, and leads to small shifts in surface gravities and
chemical abundances. \citet{frebel2013} describes a procedure to
overcome this issue, and provides a linear relation to correct the
``excitation temperatures'' derived by spectroscopy to the more
convential photometric-based temperatures. This correction is based on
data for seven metal-poor stars with metallicities, temperatures, and
gravities similar to our program stars. We apply the same procedure to
obtain our final stellar parameters. Microturbulent velocities were
determined by minimizing the trend between the abundances of \ion{Fe}{1}
lines and their reduced equivalent widths. Surface gravities were
determined from the balance of two ionization stages for iron lines
(\ion{Fe}{1} and \ion{Fe}{2}). For consistency, we allow the difference
between the abundances of the \ion{Fe}{1} and \ion{Fe}{2} lines to be in
the interval $[0.00, 0.03]$.

\begin{figure}[!ht]
\epsscale{1.20}
\plotone{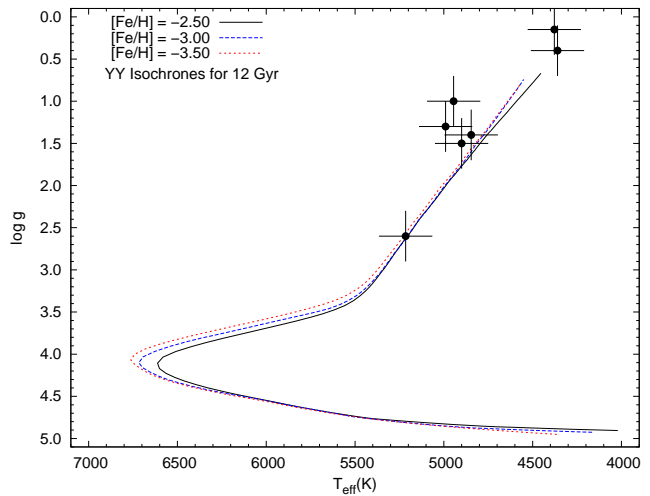}
\caption{H-R diagram for the program stars in this work. The parameters 
listed are derived from the high-resolution spectra (see Table~\ref{obstable}). 
Overplotted are the Yale-Yonsei isochrones \citep{demarque2004} 
for ages of 12 Gyr and 3 different values of [Fe/H].}
\label{isochrone}
\end{figure}

The final atmospheric parameters for our program stars are listed in
Table~\ref{obstable}. Note the good agreement between the [Fe/H] values
derived from the medium- and high-resolution analyses.
The \teff{} determinations also agree within 1-$\sigma$ for all
program stars. However, there remain
disagreements in the \logg estimates, which are greater for
the targets with \teff\ $<$ 4400~K. These temperatures are close to the
limits of the grids of medium-resolution synthetic spectra used by the
n-SSPP, which is the likely source of the discrepancies. Besides that, 
some of the features used by the n-SSPP to determine \logg\ estimates
are not present in the spectra of stars with \teff\ $<$ 4400~K.
Figure~\ref{isochrone} shows the behavior of the corrected \teff{} and
\logg{}, compared with 12~Gyr Yale-Yonsei Isochrones
\citep{demarque2004} for \metal\ = $-$3.5, $-$3.0, and $-$2.5.

\section{Abundance Analysis}
\label{abundsec}

Table~\ref{abund} lists the derived elemental
abundances (or upper limits) for 20 elements estimated from the MIKE
spectra. The $\sigma$ values listed in the tables are the standard error of the mean.
A description of the abundance analysis results is provided below.

\subsection{Techniques}

Our abundance analysis makes use of one-dimensional plane-parallel model
atmospheres with no overshooting \citep{castelli2004}, computed under
the assumption of local thermodynamic equilibrium (LTE). We use the 2011
version of the MOOG synthesis code \citep{sneden1973}, with scattering
treated with a source function that sums both absorption and scattering
components, rather than treating continuous scattering as true
absorption
\citep{sobeck2011}. 

\subsection{Abundances and Upper Limits}

Elemental abundance ratios, [X/Fe], are calculated taking solar
abundances from \citet{asplund2009}. Upper limits for elements for which
no lines could be detected can provide useful additional information for
the interpretation of the overall abundance pattern, and its possible
origin. Based on the S/N ratio in the spectral region of the line, and
employing the formula given in \citet{frebel2006b}, we derive 3$\sigma$
upper limits for Zn, Y and Eu. The abundance uncertainties, as well as
the systematic uncertainties in the abundance estimates due to the
atmospheric parameters, were treated in the same way as described in
\citet{placco2013}. Table~\ref{sys} shows how changes in each
atmospheric parameter affect the determined abundances. Also shown is
the total uncertainty for each element. In the following we present
results from the abundance analysis; discussion and interpretation of
these results is provided in Section~\ref{discuss}.

\subsubsection{Carbon and Nitrogen}

\begin{figure*}[!ht]
\epsscale{1.20}
\plotone{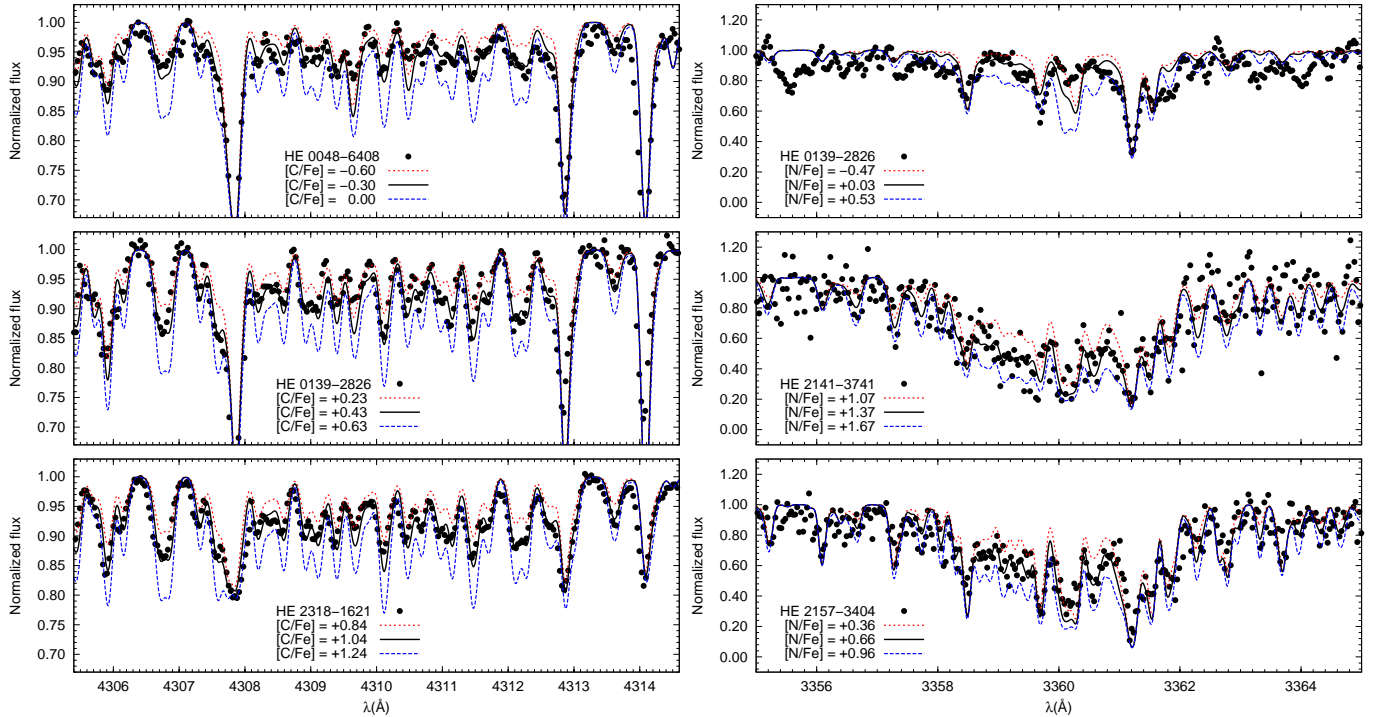}
\caption{left panels: example of a ch band used for carbon abundance determination.  
right panels: example of a nh band used for nitrogen abundance determination.  
the dots represent the observed spectra, the solid line
is the best abundance fit, and the dotted and dashed lines are the
lower and upper abundance limits, respectively, used to estimate the uncertainty.}
\label{carbon_ex}
\end{figure*}

Carbon abundances were determined from the molecular CH $G-$band feature
at 4313\, {\AA}. Data for CH molecular lines were gathered from the
compilation of \citet{frebel2007}, and references therein. For
HE~0139$-$2826, HE~2323$-$6549, and HE~2318$-$1621, which present the largest carbon
abundance ratios of our sample (\cfe\ = $+$0.48, $+$0.72, and $+$1.04,
respectively), it was possible to make a second determination using the
4322\,{\AA} region. The left panels of Figure~\ref{carbon_ex} show
examples of the carbon spectral synthesis for three of our program
stars. 

We attempted to determine nitrogen abundances from the NH band feature
at 3360\,{\AA}, which is in the lowest S/N part of the observed
high-resolution spectra. The line list for the NH feature was provided
by \citet{kurucz}, following the prescription of \citet{aoki2006}. For
HE~0139$-$2826, HE~2141$-$3741, HE~2157$-$3404, HE~2318$-$1621, and HE~2323$-$6549, the
molecular band could be synthesized, but the errors are between 0.2 and
0.5~dex, because data quality was low. The right panels of
Figure~\ref{carbon_ex} show examples of the nitrogen spectral synthesis
for three of our program stars. For HE~0048$-$6408, and HE~2233$-$4724,
we were only able to estimate a range of values for the nitrogen
abundance ratios (\abund{N}{Fe} = $[-0.28,+0.92]$ and $[-0.48,+0.72]$,
respectively). 

\subsubsection{From Na to Zn}

For all program stars, abundances for Na, Mg, Si, K, Ca, Sc, Ti, V, Cr,
Co and Ni were determined from equivalent-width analysis only (see
Table~\ref{eqw}). Aluminum abundances were determined using synthesis of
the 3944\,{\AA} line, and the derived values agree with those obtained
from equivalent-width measurements for both the 3944\,{\AA} and 
3961\,{\AA} lines. For HE~2157$-$3404, the same agreement between synthesis
and equivalent-width measurements is found for three Mn lines: 
4754\,{\AA}, 4783\,{\AA}, and 4823\,{\AA}. The Mn abundances for the other
program stars were determined via equivalent-width analysis. Two Si lines
(3906\,{\AA} and 4103\,{\AA}) were measured in each star, and their
abundances were determined via equivalent-width analysis only.

Titanium lines are found in two different ionization stages in the MIKE
spectra. The agreement between the titanium abundances derived from these
lines can also be used to assess the value of the surface gravity for
the star. The differences between \ion{Ti}{1} and \ion{Ti}{2} abundances
are within 0.12~dex for five of our program stars. There are larger
differences for two program stars: HE~0048$-$6408 (0.29~dex) and
HE~2141$-$3741 (0.25~dex). These could be related to the small number of
measured lines for \ion{Ti}{1} (5 and 3 lines, respectively). For Zn,
two lines were measured: 4722\,{\AA} and 4810\,{\AA}. For
HE~2141$-$3741 and HE~2323$-$6549, only upper limits on the Zn abundance
could be determined.

\subsubsection{Neutron-Capture Elements}

Abundances for neutron-capture elements were calculated via spectral
synthesis. For weak spectral features, upper limits were determined
following the procedure described in 
\citet{frebel2006b}.

\paragraph{Strontium, Yttrium} 

\begin{figure}[!ht]
\epsscale{1.20}
\plotone{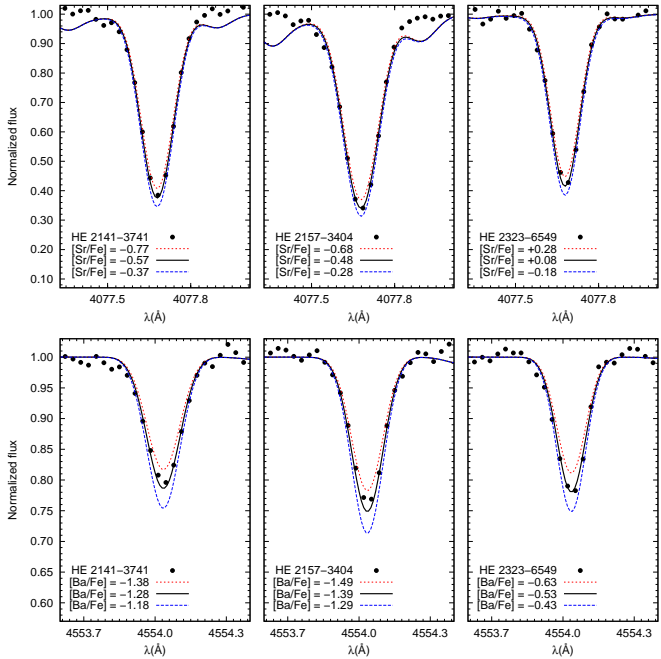}
\caption{Spectral synthesis of the Sr 4077\,{\AA} (upper
panels) and Ba 4554\,{\AA} (lower panels) lines, for three
program stars. The dots represent the observed spectra, the solid line
is the best abundance fit, and the dotted and dashed line are the lower
and upper abundance limits, indicating the abundance uncertainty.}
\label{ba_ex}
\end{figure}

The abundances of these elements are mostly determined from absorption
lines in the blue spectral regions. The Sr 4077\,{\AA} and
4215\,{\AA} lines were measurable in all program stars. In the
case of Y, abundances were derived from the 3774\,{\AA} line
for five program stars; only an upper limit on the Y abundances was
obtained for HE~2323$-$6549. No measurements or upper limits were
made for HE~2318$-$1621. In addition, the 3950\,{\AA} line
was used for upper-limit determinations of two program stars. The
upper panels of Figure~\ref{ba_ex} show the spectral synthesis of the Sr
4077\,{\AA} for three of the program stars.

\paragraph{Barium} 

Ba is mainly produced by the $s$-process in solar-system material, but
it is mostly produced by the $r$-process at low metallicity
\citep{sneden2008}. It is a useful species, as it possesses lines of
sufficient strength to be detected even in stars with
\metal~$\lesssim-3.5$. Figure~\ref{ba_ex} shows the spectral synthesis
for the 4554\,{\AA} line. In addition to this feature, the 4934\,{\AA}
was also used for abundance determinations for all program stars.

\paragraph{Europium} 

Eu is often taken as a reference element for the neutron-capture
elements that were mainly produced by the $r$-process in solar-system
material, but it can also be produced by the $s$-process in asymptotic
giant branch (AGB) nucleosynthesis. We were not able to determine
abundances for Eu, due to the non-detection of the spectral features.
Therefore, upper limits were determined from the 4129\,{\AA} and 4205\,
{\AA} lines.

\begin{figure*}[!ht]
\epsscale{1.20}
\plotone{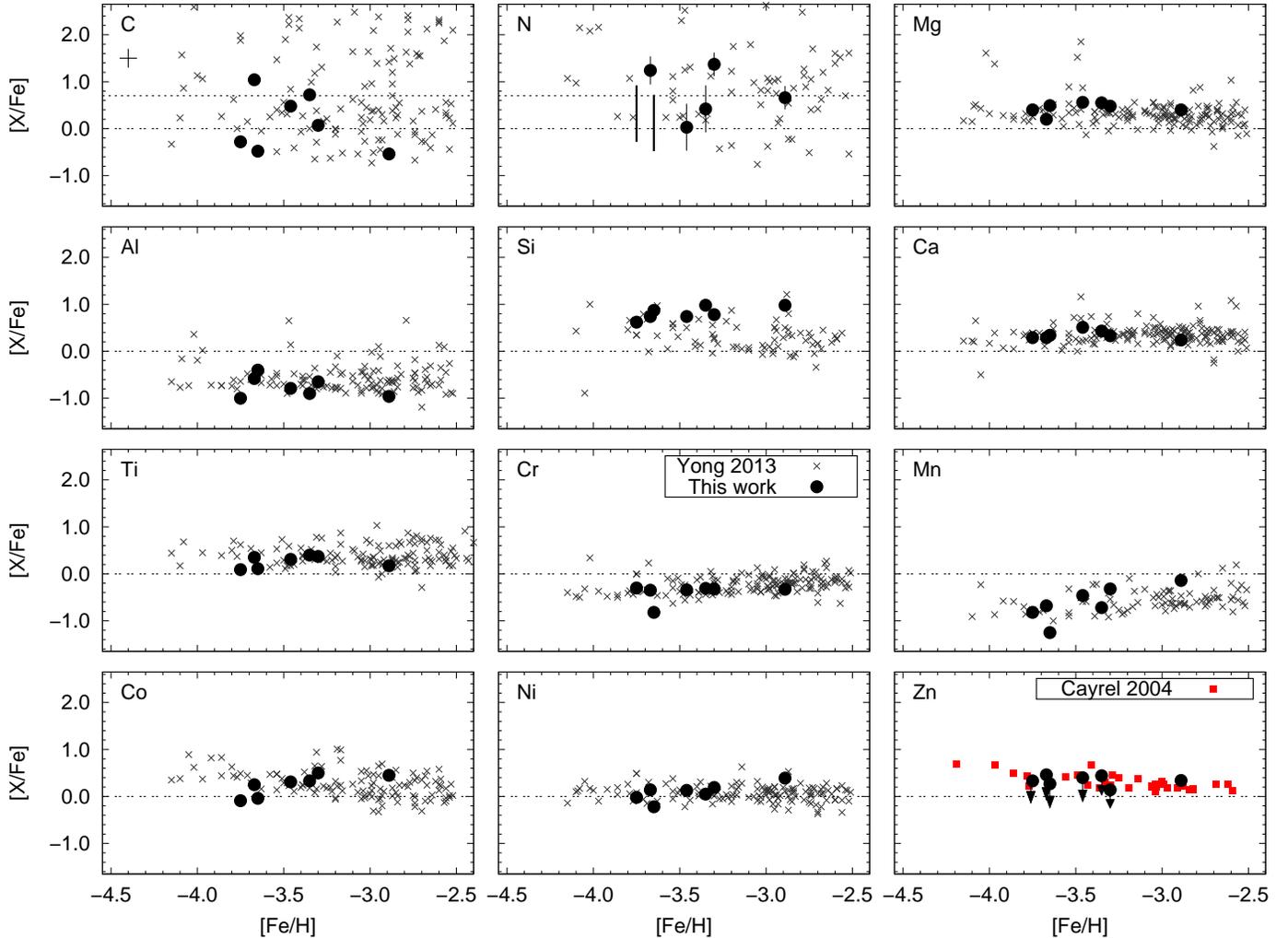}
\caption{Abundance ratios versus [Fe/H] for carbon,
nitrogen, light elements and iron-peak elements, for stars with
\metal~$<-$2.5. The circles represent the stars observed in this work,
while the squares are data taken from \citet{cayrel2004}, and crosses
are data from \citet{yong2013}. Arrows represent upper limits. Typical
error bars are shown in the top left panel.}
\label{cayrel_eps}
\end{figure*}

\section{Comparison with Other Studies and Interpretations}
\label{discuss}

Observation of new stars in the range of metallicity \metal~$\le -3.5$ is
of importance to obtain a proper understanding of the low-metallicity
tail of the Galactic MDF. The addition of HE~0048$-$6408
(\metal~=~$-$3.75), HE~0139-2826 (\metal~=~ $-$3.46), HE~2233$-$4724
(\metal~=~$-$3.65), and HE~2318$-$1621 (\metal~=~$-$3.67) to the current
number of EMP stars presents new evidence that the cutoff of the MDF at
\metal~=~$-$3.6 \citep{schorck2009,li2010} may be due to selection biases. Indeed, the
recent work of \citet{yong2013b} presents 10 new stars with
\metal~$<-$3.5, and the tail of the low-metallicity MDF when these data
are included exhibits a smooth decrease down to
\metal~=~$-$4.1. As pointed out by these authors, these data provide a
new upper limit on the MDF cutoff, and further observations in the
\metal~$<-$4 regime are required to address whether or not this feature
is a real cutoff or simply limited by small-number statistics.

We also compared the observed chemical abundances of our program stars
with data from high-resolution studies of metal-poor stars found in the
literature. This was done to identify possible deviations from the
general chemical abundance trends, and also to test different approaches
for describing these differences, if any. Figure~\ref{cayrel_eps} shows
the distribution of carbon, nitrogen, and other light-element and
iron-peak abundances as a funtion of the metallicity, [Fe/H], compared
to the stars with \metal~$<-$2.5 from \citet{cayrel2004} and
\citet{yong2013}. There are no significant differences in the abundance
trends between the program stars and the objects from the literature.

\subsection{CEMP Dependency on the Luminosity}

As can be seen from Figure ~\ref{cayrel_eps}, C and N present a larger
scatter than the remaining elements, which is believed to be due to
different formation scenarios. In particular, the CEMP stars are the
subject of a number of recent studies, which have attemped to determine
their abundance patterns and reproduce the main features by comparison
with theoretical models \citep[e.g.,][among others]{bisterzo2009,
masseron2010, bisterzo2012}. HE~2318$-$1621, with \cfe$\ = +$1.04, is clearly
a CEMP star. Besides that, one star in the upper left-hand panel of
Figure~\ref{cayrel_eps}, HE~2323$-$6549, is right above the \cfe\ = $+$0.7
line, indicating that it might be a CEMP star. However, this
classification does not take into account evolutionary mixing effects,
which can modify the initial C (and similarly, N and O) abundances of
the surface of the star. 

\begin{figure}[!ht]
\epsscale{1.20}
\plotone{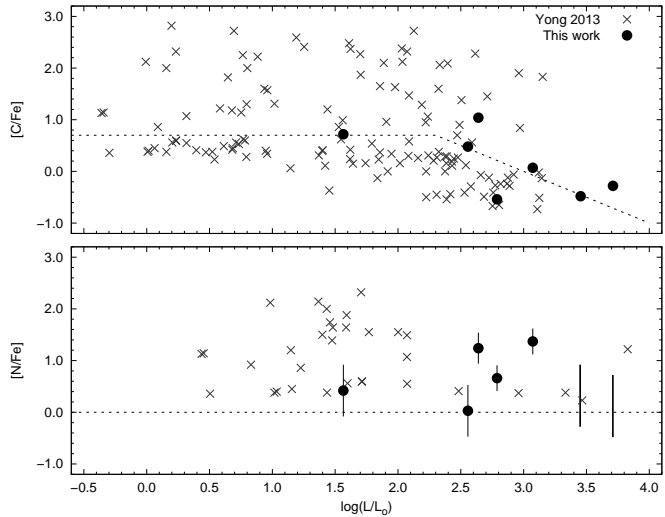}
\caption{Upper panel: Carbon abundances as a function of luminosity. The dashed line
represents the CEMP classification of \citet{aoki2007}. 
Lower panel: Nitrogen abundances as a function of luminosity. The
circles represent the stars observed in this work, and crosses are data
from \citet{yong2013}. The solid vertical lines for two star lines label the ranges
in [N/Fe] allowed by their low-S/N spectra.  The dashed line represents
the solar [N/Fe] ratio.}
\label{cfe_lum}
\end{figure}

\citet{aoki2007} suggested a CEMP classification scheme that is dependent on the
luminosity of the star, in order to take into account mixing effects.
Figure~\ref{cfe_lum} shows the behavior of the C (upper panel) and N
(lower panel) abundances as a function of the luminosity for our program
stars, as well as for the data from \citet{yong2013}. Luminosities were
estimated following the prescription of \citet{aoki2007}, and assuming
$M=0.8M_{\odot}$, typical of halo stars. With the exception of
HE~2323$-$6549, all program stars are on the upper red-giant branch, and
hence have likely suffered C-abundance depletion due to evolutionary
mixing. If the luminosity criterion of \citet{aoki2007} is adopted, six
of our program stars (all with \metal~$ < -$3.3) could have been born as
CEMP stars. 

We have assumed that the carbon-abundance depletion trend can be
extrapolated beyond $\log L/L_{\odot} = 3.0$. Since three of our stars
have luminosities larger than $\log L/L_{\odot} = 3.0$, our conclusion
on the fraction of CEMP-no stars would be affected by this assumption.
If the depletion trend levels off beyond $\log L/L_{\odot} = 3.0$, our
CEMP-no fraction would be reduced by 50\%. Additional high-resolution
spectroscopic studies of high-luminosity giants should help clarify this
issue.

According to the \citet{aoki2007} classification, a star with $\log
L/L_{\odot} = 3.5$ would be classified as CEMP if \cfe$\geq-$0.5,
implying a depletion of 1.2~dex of C during its evolution up the giant
branch. The strong depletion of C should be accompanied by an increase
in the abundance of N this star. However, evaluation of the N
abundances for our program stars on the red-giant branch indicates that
only HE~2141$-$3741 has a sufficiently high value
(\abund{N}{Fe}~$>+$1.0) to be consistent with strong C depletion. This
might be an indication that, even within the CEMP classification of
\citet{aoki2007}, the program stars above the CEMP line at high
luminosities and with low N could have had either less C than expected
(\cfe~$<+$0.7) during their main-sequence evolution, or have depleted
more C than canonically expected. In addition, the low N abundance could
also indicate lack of internal mixing in these objects, and not
constrain the initial carbon abundance. This behavior must be assessed
with higher S/N spectra in the NH region ($\sim$3360\,{\AA}) for the
three most metal-poor stars in our sample, in order to compare their
behavior with theoretical models
\citep[e.g.,][]{stancliffe2009b}.


\subsection{The ``Forbidden Zone'' and Available Fragmentation
Scenarios}

\begin{figure}[!ht]
\epsscale{1.20}
\plotone{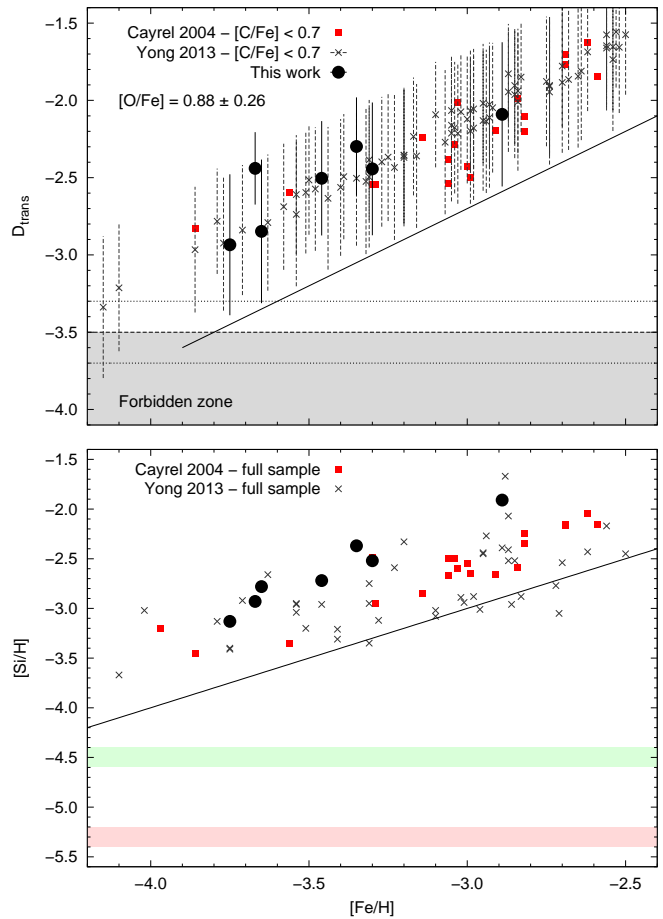}
\caption{Upper panel: D$_{trans}$ versus [Fe/H] for the observed targets and stars
from the literature with \metal~$<-$2.5. The solid line shows the scaled
solar abundance pattern, the dashed line shows the limit of D$_{trans}$
based on the model described in \citet{frebel2007b}, and the dotted
lines show the uncertainty of the model. The shaded area is the
Forbidden Zone, where there is insufficient C and O
induced cooling for low-mass star formation. Lower panel: Silicon
abundances as a function of \metal. The solid line represents
\abund{Si}{Fe}=0 as a reference. The shaded areas mark the limits of the
models presented in Figure 5 of \citet{ji2013}.}
\label{transition}
\end{figure}

Chemical abundances in EMP stars can also be used to place constraints
on the critical metallicity of the interstellar medium (ISM) for early
star formation. \citet{frebel2007b} suggested a criteria for the
formation of the first low-mass stars (``transition discriminant'' --
D$_{trans}$), using the \abund{C}{H} and \abund{O}{H} abundance ratios
in metal-poor stars. These elements, when present in the ISM, act as
efficient cooling agents, allowing gas clouds to reach temperatures and
densities that enable fragmentation to a point where a low-mass star,
still observable today, could be formed. Hence, there should be a
threshould value (D$_{trans}=-$3.5) below which C and O
fine-structure line cooling would not be able to drive low-mass star
formation \citep[``Forbidden Zone'' -- see also][for more
details]{frebel2011}. Figure~\ref{transition} (upper panel) shows the
behavior of D$_{trans}$ as a function of the metallicity, [Fe/H], for
our program stars and stars with \metal~$<-$2.5 and \cfe~$<+$0.7, from
\citet{cayrel2004} and \citet{yong2013}. The solid line shows the
scaled solar D$_{trans}$ value, the dashed line shows the limit of
D$_{trans}$, the dotted lines show the uncertainty of the model, and the
shaded area marks the Forbidden Zone.

\begin{figure*}[!ht]
\epsscale{1.10}
\plotone{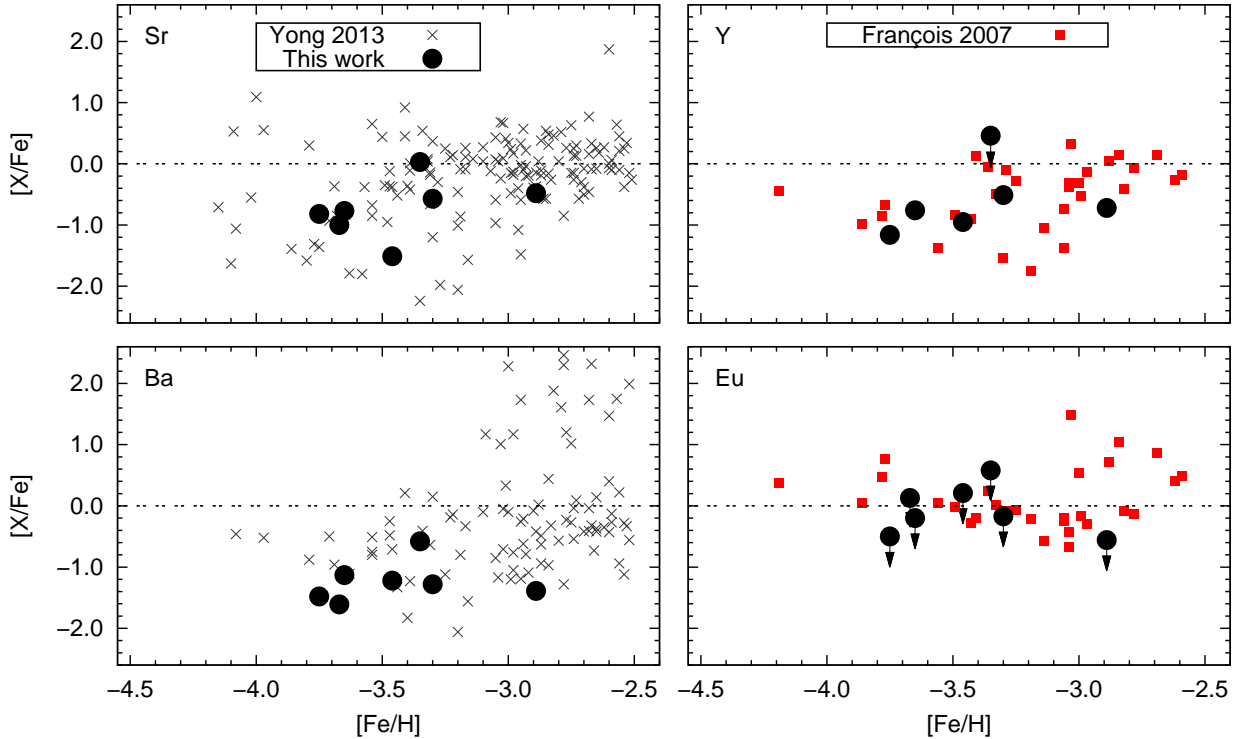}
\caption{Abundance ratios versus [Fe/H] for neutron-capture elements, 
for stars with \metal~$<-$2.5. The circles represent the program stars, the 
squares are data taken from \citet{francois2007}, and the crosses are data
from \citet{yong2013}. Arrows represent upper limits.}
\label{cayrel_nc_eps}
\end{figure*}

There are no oxygen abundances available for our program stars and the
\citet{yong2013} sample, so we estimated [O/Fe] for these targets based
on the values of \citet{cayrel2004} for stars with \cfe\ $< +0.7$
\citep[refer to Figure 8 of][lower left panel]{placco2013}. The
\abund{C}{O} ratio has a linear relation with \cfe, with an offset
of about $+0.8$~dex for the oxygen abundances. Therefore, this
assumption (\abund{O}{Fe} = 0.88 $\pm$ 0.28 for \cfe\ $< +$0.7) is a robust
estimate made to calculate D$_{trans}$. The error bars on the upper
panel of Figure~\ref{transition} account for a 2$\sigma$ difference in
\abund{O}{Fe}. It is possible to see that, for the program stars with
\metal~$<-$3.5, the D$_{trans}$ values are above the scaled solar
abundance pattern and are, within the error bars, above the Forbidden
Zone, which is consistent with the cooling scenario described by
\citet{frebel2007b}.

\citet{ji2013} consider an alternative fragmentation mode, the impact of 
thermal cooling from silicon-based dust on the formation of low-mass
stars in the early universe. This channel could account for the
formation of low-mass EMP stars in cases where an insufficient amount of
C and O were present to induce fragmentation of the gas cloud. In fact,
stars in the metallicity range \metal\ $< -$4 appear to exibit either high
Si and low C abundances, or high C and low Si abundances. The lower panel of
Figure~\ref{transition} shows the behavior of the \abund{Si}{H}
abundance as a function of the metallicity for the program stars and
stars from the literature. The solid line represents \abund{Si}{Fe} = 0,
and the shaded areas show the lower limits on the Si abundances that
would allow fragmentation \citep[see Figure 5 of][for further details on
the models]{ji2013}. One can see that our program stars, as well as the
stars from \citet{cayrel2004} and \citet{yong2013}, lie above the limits
using both criteria, so their formation can be explained by either one of these
processes. It is important to note that these stars are not sufficiently
metal poor to be used as upper limits to test these models, since they
were likely formed from gas enriched by multiple SNe.


Assuming the formation of EMP stars occured according to one of the
criteria described above, we now examine the enrichment scenarios
that would yield the required early metals prior to the formation of the
stars. Figure~\ref{cayrel_nc_eps} shows the distribution of selected
neutron-capture element abundances as a function of the metallicity for
our program stars, compared to the data from \citet{francois2007} and
\citet{yong2013}, for stars with \metal~$<-$2.5. All program
stars appear to follow the trends presented by other studies. The Ba
abundances of the stars (\abund{Ba}{Fe} $<$ 0.0), along with their C
abundances (with exception of HE~2157$-$3404), suggest that these could
be classified as CEMP-no stars \citep[see][and Figure~\ref{cfe_lum}
above]{beers2005}.

While it is true that the six EMP stars presented in this work meet the CEMP 
criteria from \citet{aoki2007}, this is based on C abundances and estimated
luminosities alone. We recognize that with such low derived C
abundances, in order to be confident of their classification as CEMP
stars, more accurate N abundances are needed. This would enable us to
justify that substantial mixing with the products of CN-processing has
indeed occurred in these stars, resulting in the transformation of C to
N. Nevertheless, for completeness, below we discuss the likely formation
scenarios for CEMP-no stars. 

In this context, the gas clouds from which the stars were formed would
have to contain C over-abundances, but lack substantial amounts of
neutron-capture elements in their composition. Two models that can
account for this abundance pattern are the first-generation,
rapidly-evolving core-collapse SN of \citet{nomoto2006} and the massive,
rapidly rotating, MMP stars of \citet{meynet2006}. Even though
both models qualitatively account for the carbon and light-element
abundance patterns found in EMP stars, they predict rather different
abundances of N, which could explain the increasing scatter of
\abund{N}{Fe} at lower metallicities. 


\subsection{The \abund{Sr}{Ba} Abundance Ratios}

We can also make use of abundance ratios such as \abund{Sr}{Ba} to
assess the likely nucleosynthesis pathways of the progenitors of our
program stars. It has been argued that elements may be formed in
different astrophysical sites, since they require different neutron
fluxes for their formation \citep{qian2003}. However, the recent study
by \citet{aoki2013} suggests that both elements are produced in the same
event (e.g., SNe II), but their observed ratio depends on specific
features of the progenitor, such as the collapse time of the star into a
black hole. In this view, the observed \abund{Sr}{Ba} ratios could be
explained by the operation of a $tr$-process \citep[``truncated
$r$-process'' -- see][for further details]{boyd2012} at early times in the
universe. 

\begin{figure}[!ht]
\epsscale{1.20}
\plotone{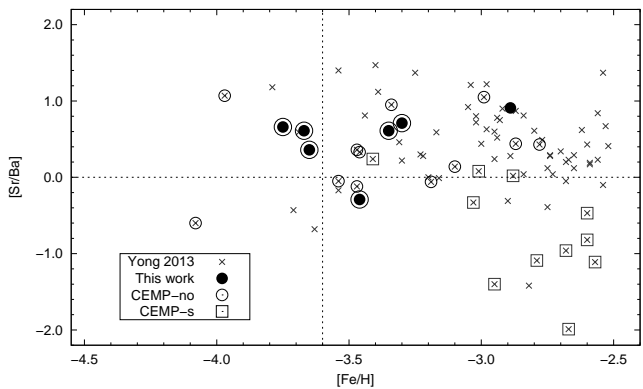}
\caption{\abund{Sr}{Ba} abundance ratio versus [Fe/H],
for stars with \metal~$<-$2.5. The filled circles represent the stars
observed in this work, the filled squares are data taken from
\citet{francois2007}, and crosses are data from \citet{yong2013}. Open
circles and squares around the symbols mark, respectively, the CEMP-no
and CEMP-$s$ stars of each sample.}
\label{sr_ba}
\end{figure}

Figure~\ref{sr_ba} shows the \abund{Sr}{Ba} abundance ratio for the
program stars, as well as data from \citet{francois2007} and from
\citet{yong2013}. Also shown are the CEMP-$s$ (\cfe\ $ > +$0.7 and
\abund{Ba}{Fe} $> +$1.0) and CEMP-no (\cfe\ $> +$0.7 and \abund{Ba}{Fe} $<$ 0.0) 
stars identified in each sample. The presence of low \abund{Sr}{Ba}
ratios at \metal~$ > -3.0$ is expected, since this metallicity regime is
thought to reflect the onset of the $s$-process in the Galaxy
\citep[see][for further details]{simmerer2004,sivarani2004}, 
and Ba is mainly formed by the $s$-process in solar-system material
\citep[85\% according to][]{burris2000}. Moreover, the majority of these
stars are CEMP-$s$ (with \abund{Ba}{Fe}~$>+$1.0), whose abundance patterns
are a direct evidence of mass transfer episodes from a low-mass AGB star
in a binary system \citep[see][for further details]{placco2013}.

In the \metal~$<-3.0$ regime, the formation of Sr and Ba is likely
dominated by the $r$-process \citep[or alternatively by the $tr$-process, as
mentioned above, or else by an $s$-process associated with the
explosions of massive stars; see][]{cescutti2013}. There appears to
exist increasing scatter for \abund{Sr}{Ba} ratios with decreasing
metallicity, which is consistent with the behavior of the data analyzed
by \citet{aoki2013}. However, a number of stars exhibit \abund{Sr}{Ba}
ratios that are not entirely consistent with the cutoff at
\abund{Sr}{Ba} $>$ 0.0 suggested by \citet{aoki2013}. 

\section{Conclusions}
\label{final}

In this work we analyzed seven newly discovered VMP/EMP stars, originally
selected from the low-resolution HES plates, and followed up with
medium-resolution, then high-resolution spectroscopy. According to our
analysis of the high-resolution spectra, six of them exhibit
\metal~$< -$3.0, and four have \metal~$\le -$3.5.

We have analyzed selected light and neutron-capture elements for the
program stars; there are no significant abundance differences when
compared with results from the literature. Six of our targets are
consistent with the CEMP-no classification of \citet{aoki2007},
and exhibit enhancements in C (corrected for evolutionary mixing)
and a lack of neutron-capture elements (\abund{Ba}{Fe} $<$ 0.0). However, more
accurate N abundances are needed in order to confirm this. The
abundances in the sample are also consistent with C-normal EMP stars
from the literature, and can also be used to study the yields of the
early stellar generations to form in the Galaxy.

It is worth emphasizing that the \metal~$<-$3.5 metallicity regime is of
particular importance in determining the true shape of the MDF, and
further observations of EMP stars can help understand whether the
observed MDF tail at \metal\ = $-$4.1 is real or a result of an incomplete
inventory of low-metallicity stars.

\acknowledgments 

VMP acknowledges FAPESP (2012/{}13722-1) funding. TCB
acknowledges partial support for this work from grants PHY 02-15783 and PHY 08-22648; 
Physics Frontier Center/{}Joint Institute or
Nuclear Astrophysics (JINA), awarded by the U.S. National Science
Foundation. NC is was supported by Sonderforschungsbereich SFB 881
``The Milky Way System'' (subprojects A4 and A5) of the German Research
Foundation (DFG). YSL is a Tombaugh Fellow. CRK acknowledges
support from Australian Research Council (Super Science Fellowship;
FS110200016). SR and RMS thank FAPESP, Capes, and CNPq for
partial financial support.

\clearpage
\ltab

\begin{center}
\begin{deluxetable}{lrrrrrrr}
\rotate
\tabletypesize{\scriptsize}
\tablecaption{Observational Data for the Observed Candidates \label{candlist}}
\tablehead{
\colhead{} & 
\colhead{HE~0048$-$6408} &
\colhead{HE~0139$-$2826} &
\colhead{HE~2141$-$3741} &
\colhead{HE~2157$-$3404} &
\colhead{HE~2233$-$4724} &
\colhead{HE~2318$-$1621} &
\colhead{HE~2323$-$6549}}
\tablewidth{0pt}
\startdata
$\alpha$ (J2000)  &    00:50:45.3 &    01:41:36.8 &    21:44:51.1 &    22:00:17.6 &    22:35:59.2 &    23:21:21.5 &    23:26:14.5 \\
$\delta$ (J2000)  & $-$63:51:50.0 & $-$28:11:03.0 & $-$37:27:55.0 & $-$33:50:04.0 & $-$47:08:36.0 & $-$16:05:06.0 & $-$65:33:06.0 \\
V (mag)           &          12.9 &          13.7 &          13.4 &          13.2 &          14.8 &          13.1 &          14.3 \\
\jk               &          0.63 &          0.53 &          0.51 &          0.53 &          0.64 &          0.51 &          0.50 \\
KP (\,{\AA})      &           4.1 &           4.0 &           2.8 &           4.9 &           6.6 &           3.7 &           4.5 \\
\hline
\multicolumn{8}{c}{Medium Resolution} \\
\hline
Date              &    2011 10 17 &    2011 10 17 &    2011 08 03 &    2011 08 03 &    2011 10 17 &    2011 10 17 &    2011 08 03 \\
UT                &      04:27:26 &      06:09:16 &      03:33:00 &      04:11:00 &      01:16:05 &      02:57:46 &      08:29:00 \\
\enddata
\end{deluxetable}
\end{center}


\clearpage 

\begin{deluxetable}{@{}l@{}r@{}r@{}r@{}r@{}r@{}r@{}r@{}r@{}r@{}r@{}r@{}r@{}r@{}r@{}r@{}r@{}r@{}} 
\tabletypesize{\tiny}
\rotate
\tablecolumns{5} 
\tablewidth{0pc} 
\tablecaption{\label{eqw} Equivalent Width Measurements} 
\tablehead{ 
\colhead{}&
\colhead{}&
\colhead{}&
\colhead{}&
\multicolumn{2}{c}{HE0048$-$6408}&
\multicolumn{2}{c}{HE0139$-$2826}&
\multicolumn{2}{c}{HE2141$-$3741}&
\multicolumn{2}{c}{HE2157$-$3404}&
\multicolumn{2}{c}{HE2233$-$4724}&
\multicolumn{2}{c}{HE2318$-$1621}&
\multicolumn{2}{c}{HE2323$-$6549}\\
\cline{5-6} \cline{7-8} \cline{9-10} \cline{11-12} \cline{13-14} \cline{15-16} \cline{17-18}\\ 
\colhead{Ion}& 
\colhead{$\lambda$}& 
\colhead{$\chi$} &  
\colhead{$\log\,gf$}& 
\colhead{$W$}&
\colhead{$\log\epsilon$}&
\colhead{$W$}&
\colhead{$\log\epsilon$}&
\colhead{$W$}&
\colhead{$\log\epsilon$}&
\colhead{$W$}&
\colhead{$\log\epsilon$}&
\colhead{$W$}&
\colhead{$\log\epsilon$}&
\colhead{$W$}&
\colhead{$\log\epsilon$}&
\colhead{$W$}&
\colhead{$\log\epsilon$}\\
\colhead{}& 
\colhead{{\AA}}& 
\colhead{eV} &  
\colhead{}& 
\colhead{m{\AA}}&
\colhead{}&
\colhead{m{\AA}}&
\colhead{}&
\colhead{m{\AA}}&
\colhead{}&
\colhead{m{\AA}}&
\colhead{}&
\colhead{m{\AA}}&
\colhead{}&
\colhead{m{\AA}}&
\colhead{}&
\colhead{m{\AA}}&
\colhead{}} 
\startdata 
C CH  & 4313.00   & \nodata & \nodata   & syn	  & 4.4     & syn     &   5.4	& syn	  &   5.2   & syn     &   5.0	& syn	  & 4.3\xx  &     syn &    5.30 & syn	  &   5.8   \\
C CH  & 4322.00   & \nodata & \nodata   & \nodata & \nodata & syn     &   5.5	& \nodata & \nodata & \nodata & \nodata & \nodata & \nodata &     syn &    5.30 & syn	  &   5.8   \\
N NH  & 3360.00   & \nodata & \nodata   & \nodata & \nodata & syn     &   4.4	& syn	  &   5.9   & syn     &   5.6	& \nodata & \nodata &     syn &    5.40 & syn	  &   4.9   \\
Na I  & 5889.95   & 0.00    &    0.11   & 134.7   &   2.6   & 121.4   &   3.3	& 127.6   &   3.3   & 165.9   &   4.0	& 192.8   &   3.3   &     0.1 &    0.00 & 126.6   &   3.6   \\
Na I  & 5895.92   & 0.00    & $-$0.19   & 102.7   &   2.5   & 100.1   &   3.2	&  98.9   &   3.1   & 131.0   &   3.7	& 167.1   &   3.2   &   108.7 &    3.28 &  94.0   &   3.3   \\
Mg I  & 3829.36   & 2.71    & $-$0.21   & \nodata & \nodata & 130.0   &   4.8	& 128.7   &   4.7   & \nodata & \nodata & \nodata & \nodata &    96.9 &    3.94 & 125.7   &   4.7   \\
Mg I  & 3832.30   & 2.71    &    0.27   & \nodata & \nodata & 150.1   &   4.6	& \nodata & \nodata & \nodata & \nodata & \nodata & \nodata & \nodata & \nodata & \nodata & \nodata \\
Mg I  & 3838.29   & 2.72    &    0.49   & \nodata & \nodata & 169.5   &   4.6	& \nodata & \nodata & \nodata & \nodata & \nodata & \nodata &   130.6 &    4.02 & \nodata & \nodata \\
\enddata 
\tablenotetext{a}{Upper limits.}
\end{deluxetable}

\clearpage

\begin{center}
\begin{deluxetable}{ccccccccc}
\tabletypesize{\scriptsize}
\tablecolumns{9}
\tablewidth{0pc}
\tablecaption{\label{obstable} Derived Stellar Parameters}
\tablehead{&
\multicolumn{3}{c}{Medium Resolution}&&
\multicolumn{4}{c}{High Resolution}\\
\cline{2-4} \cline{6-9}\\ 
\colhead{}&
\colhead{\teff{}(K)    }&
\colhead{\logg{}(cgs)  }&
\colhead{\metal        }&
\colhead{              }&
\colhead{\teff{}(K)    }&
\colhead{\logg{}(cgs)  }&
\colhead{$\xi$(km/s)   }&
\colhead{\metal{}      }}
\startdata
HE~0048$-$6408 & 4524 (125) & 1.73 (0.35) &  $-$3.83 (0.20) && 4378 (150) & 0.15 (0.30) & 2.85 (0.30) & $-$3.75 (0.10) \\
HE~0139$-$2826 & 5011 (125) & 1.42 (0.35) &  $-$3.50 (0.20) && 4900 (150) & 1.50 (0.30) & 1.70 (0.30) & $-$3.46 (0.10) \\
HE~2141$-$3741 & 5075 (125) & 1.33 (0.35) &  $-$3.04 (0.20) && 4945 (150) & 1.00 (0.30) & 2.00 (0.30) & $-$3.30 (0.10) \\
HE~2157$-$3404 & 5027 (125) & 1.71 (0.35) &  $-$2.91 (0.20) && 4990 (150) & 1.30 (0.30) & 2.00 (0.30) & $-$2.89 (0.10) \\
HE~2233$-$4724 & 4587 (125) & 1.26 (0.35) &  $-$3.43 (0.20) && 4360 (150) & 0.40 (0.30) & 3.00 (0.30) & $-$3.65 (0.10) \\
HE~2318$-$1621 & 5043 (125) & 1.58 (0.35) &  $-$3.75 (0.20) && 4846 (150) & 1.40 (0.30) & 1.75 (0.30) & $-$3.67 (0.10) \\
HE~2323$-$6549 & 5317 (125) & 2.30 (0.35) &  $-$3.35 (0.20) && 5215 (150) & 2.60 (0.30) & 1.55 (0.30) & $-$3.35 (0.10) 
\enddata
\end{deluxetable}
\end{center}

\clearpage

\begin{deluxetable}{@{}l@{}rrrrr@{}rrrrr@{}rrrrrrrrr@{}}
\tablewidth{0pt}  
\tabletypesize{\tiny}
\rotate
\tablecaption{Abundances for Individual Species \label{abund}}
\tablehead{}
\startdata 
& \multicolumn{4}{c}{HE~0048$-$6408} &  & \multicolumn{4}{c}{HE~0139$-$2826} &  & \multicolumn{4}{c}{HE~2141$-$3741} & & \multicolumn{4}{c}{HE~2157$-$3404} \\
 \cline{2-5} \cline{7-10}\cline{12-15}\cline{17-20}

& $\log\epsilon$\,(X) & $\mbox{[X/Fe]}$ & $\sigma$ & $N$ & & $\log\epsilon$\,(X) & $\mbox{[X/Fe]}$ & $\sigma$ & $N$ 
      & & $\log\epsilon$\,(X) & $\mbox{[X/Fe]}$ & $\sigma$ & $N$ & & $\log\epsilon$\,(X) & $\mbox{[X/Fe]}$ & $\sigma$ & $N$ \\
\hline

C      &      4.40 &  $-$0.28  &     0.25  &	    1 & &     5.45  &	  0.48  &    0.25  &       2 & &     5.20  &     0.07  &    0.30  &       1 & &	   5.00  &  $-$0.54  &    0.30  &       1  \\	  
N      &   \nodata &   \nodata &   \nodata &  \nodata & &     4.40  &	  0.03  &    0.50  &       1 & &     5.90  &     1.37  &    0.25  &       1 & &	   5.60  &     0.66  &    0.25  &       1  \\	  
Na  I  &     2.55  &	 0.06  &     0.09  &	    2 & &     3.24  &	  0.46  &    0.08  &       2 & &     3.23  &     0.29  &    0.14  &       2 & &	   3.88  &     0.53  &    0.20  &       2  \\
Mg  I  &     4.25  &	 0.40  &     0.03  &	    5 & &     4.70  &	  0.56  &    0.03  &       7 & &     4.78  &     0.48  &    0.06  &       5 & &	   5.11  &     0.40  &    0.02  &       4  \\
Al  I  &     1.70  &  $-$1.00  &     0.15  &	    1 & &     2.29  &  $-$0.79  &    0.20  &       1 & &     2.50  &  $-$0.65  &    0.10  &       1 & &	   2.60  &  $-$0.96  &    0.15  &       1  \\		
Si  I  &     4.38  &	 0.62  &     0.04  &	    2 & &     4.79  &	  0.74  &    0.10  &       1 & &     4.99  &     0.78  &    0.10  &       1 & &	   5.60  &     0.98  &    0.10  &       1  \\
\enddata              
\tablenotetext{a}{$\mbox{[FeI/H]}$ and $\mbox{[FeII/H]}$ values}
\end{deluxetable}

\clearpage

\begin{deluxetable}{lrrrr}
\tabletypesize{\scriptsize}
\tablewidth{0pc}
\tablecaption{Example Systematic Abundance Uncertainties for HE~2157$-$3404 \label{sys}}
\tablehead{
\colhead{Elem}&
\colhead{$\Delta$\teff}&
\colhead{$\Delta$\logg}&
\colhead{$\Delta\xi$}&
\colhead{$\sigma_{\rm tot}$}\\
\colhead{}&
\colhead{$+$150\,K}&
\colhead{$+$0.5 dex}&
\colhead{$+$0.3 km/s}&
\colhead{}}
\startdata
CH    & $-$0.30 &    0.20 & 0.00 & 0.36 \\
Na I  & $-$0.17 &    0.08 & 0.18 & 0.26 \\
Mg I  & $-$0.08 &    0.03 & 0.03 & 0.09 \\
Al I  & $-$0.14 &    0.01 & 0.18 & 0.23 \\
Si I  & $-$0.16 &    0.06 & 0.10 & 0.19 \\
Ca I  & $-$0.11 &    0.03 & 0.02 & 0.11 \\
Sc II & $-$0.07 & $-$0.13 & 0.07 & 0.16 \\
Ti I  & $-$0.18 &    0.05 & 0.02 & 0.19 \\
Ti II & $-$0.06 & $-$0.13 & 0.08 & 0.16 \\
Cr I  & $-$0.18 &    0.06 & 0.09 & 0.21 \\
Mn I  & $-$0.17 &    0.06 & 0.08 & 0.20 \\
Fe I  & $-$0.16 &    0.05 & 0.05 & 0.17 \\
Fe II & $-$0.01 & $-$0.15 & 0.05 & 0.16 \\
Co I  & $-$0.20 &    0.06 & 0.10 & 0.23 \\
Ni I  & $-$0.16 &    0.06 & 0.09 & 0.19 \\
Zn I  & $-$0.09 & $-$0.05 & 0.00 & 0.10 \\
Sr II & $-$0.15 & $-$0.20 & 0.20 & 0.32 \\
Ba II & $-$0.20 & $-$0.10 & 0.00 & 0.22
\enddata
\end{deluxetable}

\end{document}